\documentclass[twocolumn,letterpaper]{revtex4-1}

\usepackage{soul}

\usepackage{adjustbox}
\usepackage{color}
\usepackage{graphicx}
\usepackage{subfigure}
\usepackage{amsmath}
\usepackage{amsfonts}
\usepackage{citesort}
\usepackage{bm}

\newcommand{\comment}[1]{}

\newcommand{\overbar}[1]{\mkern 1.5mu\overline{\mkern-1.5mu#1\mkern-1.5mu}\mkern 1.5mu}

\newcommand{\imag}{\mathrm{i}}
\newcommand{\total}{\mathrm{d}}

\begin{document}

\title{Fundamental issues with light propagation through $\mathcal{PT}$-symmetric systems}

\author{F.~A.~Shuklin$^{1\ast}$, C.~Tserkezis$^{1}$, N.~Asger~Mortensen$^{1,2}$, C.~Wolff$^{1}$}
\affiliation{
    $^1$Center for Nano Optics, University of Southern Denmark, Campusvej 55, DK-5230~Odense~M, Denmark\\
    $^2$Danish Institute for Advanced Study, University of Southern Denmark, Campusvej 55, DK-5230~Odense~M, Denmark
}

\email{fesh@mci.sdu.dk}

\date{\today}

\begin{abstract}
We analyse the emergence of unphysical superluminal group velocities in Su--Schrieffer--Heeger (SSH) parity-time ($\mathcal{PT}$) symmetric chains, and explore the origins of such a behaviour. By comparing the band structure of an infinite loss-gain SSH chain with that of a one-dimensional Bragg stack, we first exclude insufficient coupling consideration in the tight-binding description as the cause of group-velocity divergence. We then focus on material dispersion, and show that indeed, restoring causality in the description of both the lossy and the gain components resolves the problem and recovers finite group velocities, whose real part can only exceed the speed of light in vacuum when accompanied by a significant imaginary part. Our analysis introduces thus the required practical limits in the performance of common $\mathcal{PT}$-symmetric systems.
\end{abstract}

\maketitle

\section{Introduction}
\label{Sec:Intro}

The quest for reaching physics beyond 
Hermitian Hamiltonians is crucial for the theoretical description of non-conserving physical systems, e.g., open quantum systems~\cite{BreuerPetruccione:2002,Rotter:2015}.
Much work has been invested in providing a solid theoretical base to non-Hermitian physics~\cite{Ashida:2020}, which is a more recently emerging direction also in photonics~\cite{El-Ganainy:2018,Zhao:2018,Krasnok:2021,Parto:2021}. 
One of the most significant and fruitful achievements on this way was the introduction of $\mathcal{PT}$-symmetric Hamiltonians by Bender and Boettcher~\cite{BenderBoettcher:1998}.
They discovered that a non-conserving Hamiltonian $\mathcal{H}(\gamma)$, with $\gamma$ being a parameter of non-Hermitivity [i.e., such that $\mathcal{H}(\gamma)$ at $\gamma=0$ is Hermitian], can have real eigenvalues if it is invariant under the parity, $\mathcal{P}$, and time-reversal, $\mathcal{T}$, transformation $\mathcal{PT}\mathcal{H}(\gamma)(\mathcal{PT})^{-1} = \mathcal{H}(\gamma)$, which implies that the Hamiltonian and the parity-time operator share a common set of eigenstates. 
Relying on the previous condition, one can construct a $\mathcal{PT}$-symmetric Hamiltonian by imposing on its complex potential the conditions~\cite{Vinogradov:2014}
\begin{equation}\label{eq:potentialCondition}
	V(\mathbf{r},\gamma) = V^*(-\mathbf{r},\gamma),
\end{equation}
where $\mathbf{r}$ is a position vector. 

However, eigenvalues of such Hamiltonians 
cannot be purely real for any arbitrarily large parameter $\gamma$. 
When $\gamma$ exceeds some critical value the system exhibits an abrupt symmetry-breaking phase transition.  
In this regime, the Hamiltonian and $\mathcal{PT}$ operators no longer share the same eigenstates, and the eigenvalues of the system cease to be real valued~\cite{BenderBoetcherMeisinger:1999}. 
This critical value of $\gamma$ 
constitutes an exceptional point (EP), being intriguing by itself: at the EP, at least two eigenvalues of the system become degenerate, and the corresponding eigenstates coalesce, which gives rise to interesting physics and unusual phenomena~\cite{Krasnok:2021, Moiseyev:2021}. 

Originating from quantum physics, the concept of $\mathcal{PT}$ symmetry has spread to optics, first introduced in the paraxial wave equation, being formally equivalent to the Schr{\"o}dinger equation~\cite{El-Ganainy:2007, Guo:2009, Ruter:2010}.
In optical systems, $\mathcal{PT}$ symmetry can be established by incorporating gain~(G) and loss (L) so that the permittivity $\varepsilon$ plays the role of a complex potential of a quantum Hamiltonian. Considering such a system, the condition Eq.~(\ref{eq:potentialCondition}) for the complex permittivity takes the form 
\begin{subequations}
\label{eq:permittivityCondition}
\begin{align}
	\Re\{\varepsilon(\mathbf{r},\gamma)\} = \Re\{\varepsilon(-\mathbf{r},\gamma)\}, \\
\Im\{\varepsilon(\mathbf{r}, \gamma)\} = -\Im\{\varepsilon(-\mathbf{r}, \gamma)\}.
\end{align}
\end{subequations}
This setup is \emph{per se} not restricted to the paraxial approximation, as it may be applied to numerous optical problems~\cite{Klaiman:2008, El-Ganainy:2018, Vinogradov:2014,Khurgin:2021}. 
It has turned out to be a fertile ground for a new field of photonics, gaining many fundamental results and application proposals, including sensitivity enhancement~\cite{Wiersig:2014, LiuSensing:2016, Chen:2017, Hodaei:2017,Wiersig:2020}, 
$\mathcal{PT}$-symmetric lasers~\cite{LonghiPTLaser:2010, Feng:2014, HodaeiLaser:2014}, and
$\mathcal{PT}$-symmetric optical diodes based on non-reciprocal light propagation~\cite{Feng:2011, Ramezani:2010}. 
Many fundamental theoretical studies have looked into topological properties of $\mathcal{PT}$-symmetric systems~\cite{Weimann:2017, Yuce:2015, Yuce2:2015}, 
nonlinear effects~\cite{Konotop:2016a} 
and have searched for higher-order EPs~\cite{Musslimani:2008, Lin:2016, Mandal:2021}.
Another fruitful research direction is pulse propagation through $\mathcal{PT}$-symmetric systems, demonstrating effects like double refraction and non-reciprocal diffraction or Bloch oscillations~\cite{Markis:2008,Zhong:2018}. 

However, this new rapidly growing field faces also inherent difficulties. For instance, perfectly balancing gain and loss is a very challenging practical task, so experimental observations of $\mathcal{PT}$-symmetry related phenomena are quite difficult~\cite{Guo:2009}. To mitigate this, so-called ``quasi''-$\mathcal{PT}$-symmetric systems were proposed, containing only passive components~\cite{Zhong:2016,Ornigotti:2014}. Fundamental limitations, like unavoidable noise limiting sensitivity enhancement near the EP~\cite{Mortensen:2018,Wolff:2019,Langbein:2018}, create even greater difficulties.
Last, but not least, Eqs.~\eqref{eq:permittivityCondition}, which enforce
$\mathcal{PT}$ symmetry of the overall system, clash most severely with
causality when assumed to apply in a broader frequency range in conjunction with dispersive $\varepsilon$~\cite{Vinogradov:2014} --- note that the permittivity must fundamentally satisfy the high frequency limit 
$\lim_{\omega \rightarrow \infty} \varepsilon(\omega) = 1$ (where
$\omega$ is the angular frequency):
if $\varepsilon(\mathbf{r})$ at position $\mathbf{r}$ is a ``regular'' 
material with a retarded response, then $\varepsilon(-\mathbf{r})$ at the 
inverted position must concurrently exhibit an advanced response whose polarization is 
\emph{exclusively} an anticipation of future stimuli.

The main theme of this manuscript is the propagation of light in 
$\mathcal{PT}$-symmetric systems composed of coupled resonators.
This includes a variety of different geometries such as coupled dielectric
ring resonators~\cite{Hodaei:2017} or plasmonic nanoparticles~\cite{Sanders:2020}.
However, the understanding of their fundamental properties such as the existence of real EPs does not depend on the specific implementation, and is usually analysed using more abstract theoretical tools such as coupled-mode theory (CMT)~\cite{YarivCoupledmode:1973, Chien:2007, Fu:2020}, eventually leading to a tight-binding chain, which may be as short as only two sites.
However, even within a tight-binding model, there is a significant range of approximations common to Hermitian problems that can lead to unphysical behavior in non-Hermitian cases, including $\mathcal{PT}$-symmetric systems.
In order to trace the origin of such theoretical artifacts, we analyse the typical tight-binding model alongside the simplest possible realization of a coupled-resonator system in photonics: a Bragg stack.

\section{Light propagation in $\mathcal{PT}$-symmetric systems}

We start with an analysis of light propagation in a simple one-dimensional (1D) infinite $\mathcal{PT}$-symmetric system.
The corresponding tight-binding model takes the form of a simple Su--Schrieffer--Heeger (SSH) chain~\cite{Su:1979}.
It is a 1D lattice composed of two sublattices; a first one with gain and a second one with an equal amount of loss. 
Consequently, each unit cell --- being net passive --- contains one site with gain and one with loss, corresponding to the on-site terms $\omega_0 + i\gamma$ and $\omega_0 - i\gamma$, respectively.
Here, $\omega_{0}$ is the real part of the on-site angular frequency, and $\pm i\gamma$ is the gain/loss rate (the Hermitian case is recovered in the limit $\gamma\rightarrow 0$). 
We assume only nearest-neighbor interactions in the chain; sites inside a unit cell coupled with an intra-cell coupling rate $\kappa_\text{in}$, neighboring sites of different unit cells coupled with an inter-cell coupling rate $\kappa_\text{out}$.
A sketch of the system is presented at the inset of Fig.~\ref{fig:figure_1}(a). 

The dynamics of the system is governed by the equation obtained within CMT,
\begin{subequations}
\begin{align}\label{eq:sshEquation}
\imag \partial_t\textbf{a}(t) = \mathcal{H}\textbf{a}(t), 
\end{align}
where the state-vector $\textbf{a}(t)$ contains the amplitude on each lattice site, and $\mathcal{H}$ is the tridiagonal SSH Hamiltonian:
\begin{equation}\label{eq:sshHamiltonian}
    \mathcal{H} = 
    \begin{pmatrix}
    \ddots & & & \\
    &\kappa_\text{out} & \omega_0 - \imag\gamma & \kappa_\text{in}^*& 0 \\
    &0 & \kappa_\text{in} & \omega_0 + \imag\gamma & \kappa_\text{out}^* \\
    & & & & & \ddots
    \end{pmatrix}.
\end{equation}
\end{subequations}

One of the key characteristics of periodic media is the band structure (BS) --- a versatile concept originating from condensed-matter physics and the Hermitian treatment of electrons in periodic potentials~\cite{Kohn:1999,Reuter:2016} --- which, in the present case, generalizes to the mutual relation
between the angular frequency $\omega$ and the wavenumber $k$ of each state. The BSs of crystals are thus commonly understood to consist of the
real frequencies related to real-valued wave numbers, implying also intuitive ways to describe other physical observables for Hermitian systems.
The BS inherently provides information on other properties of the
wave system such as the group velocity $v_\text{group}$ in form of the BS derivative $\partial \omega / \partial k$ and the total electromagnetic density of states (DOS) as being proportional to the inverse group velocity.
However, even in solid-state physics, it has been known that this interpretation of the BS derivative is restricted to the real $k$-axis and that real eigenenergies exist, e.g., inside band gaps for non-real $k$ along the so-called Heine's lines of real energy~\cite{Heine:1963}.
At the outermost points of these lines, the different energy bands are connected via a 
branch point and here $\partial \omega/\partial k \rightarrow \infty.$
This apparent divergence of the ``group velocity'' is of no consequence, because the corresponding states are purely evanescent and do not carry any energy.

The above discussion of properties applies to periodic photonic systems provided they are Hermitian, i.e., for real-valued permittivities independent of $\omega$.
However, assuming a Hermitian system is only an approximation in photonic systems, because material dispersion (MD) and dissipation are inevitably omnipresent, being a natural consequence of the principle of causality~\cite{Dethe:2019}.
The case of weak loss and dispersion does not change anything fundamentally and can be easily treated as a perturbation, while for strong dispersion and loss the expression for the velocity of pulse propagation
requires significant modifications.
Interestingly, this is \emph{not} a result of the lack of Hermiticity, but of the
derivative $\partial \omega/\partial k$ becoming complex valued, i.e., non-real~\cite{Wolff:2018}.
It should be stressed that this derivative maintains its meaning as a group velocity if the real $k$ points are associated with real $\omega$ values, irrespective of the Hermiticity of the Hamiltonian.
Therefore, it can happen that seemingly correct band structures contain unphysical features, e.g., predicting pulses that would propagate with superluminal speeds, if strong loss or dispersion are not treated appropriately, and in consistency with the principles of causality.

A good example for illustrating this problem is the phenomenon of band back-bending observed in photonic crystals (PhCs) with strongly $\omega$-dependent constituent permittivities.
Band-structure plots computed for metallic PhCs neglecting material loss seem to predict infinite group velocities~\cite{Tserkezis:2009}. 
In fact, these systems do not constitute conventional Hermitian eigenvalue problems despite having a real band structure, e.g., because the eigenstates for a given $k$ mathematically do not form an orthogonal basis of the function space.
However, the diverging group velocity is clearly an artifact of ignoring the principles of causality and the resulting Kramers-Kronig relations, while the group velocity remains finite (albeit complex valued) once a dispersive material loss is taken into consideration.

\section{Superluminal pulse propagation}

\subsection{Infinite group velocity in tight-binding chains}

\begin{figure}[h]
    \centering
    \includegraphics[width=\columnwidth]{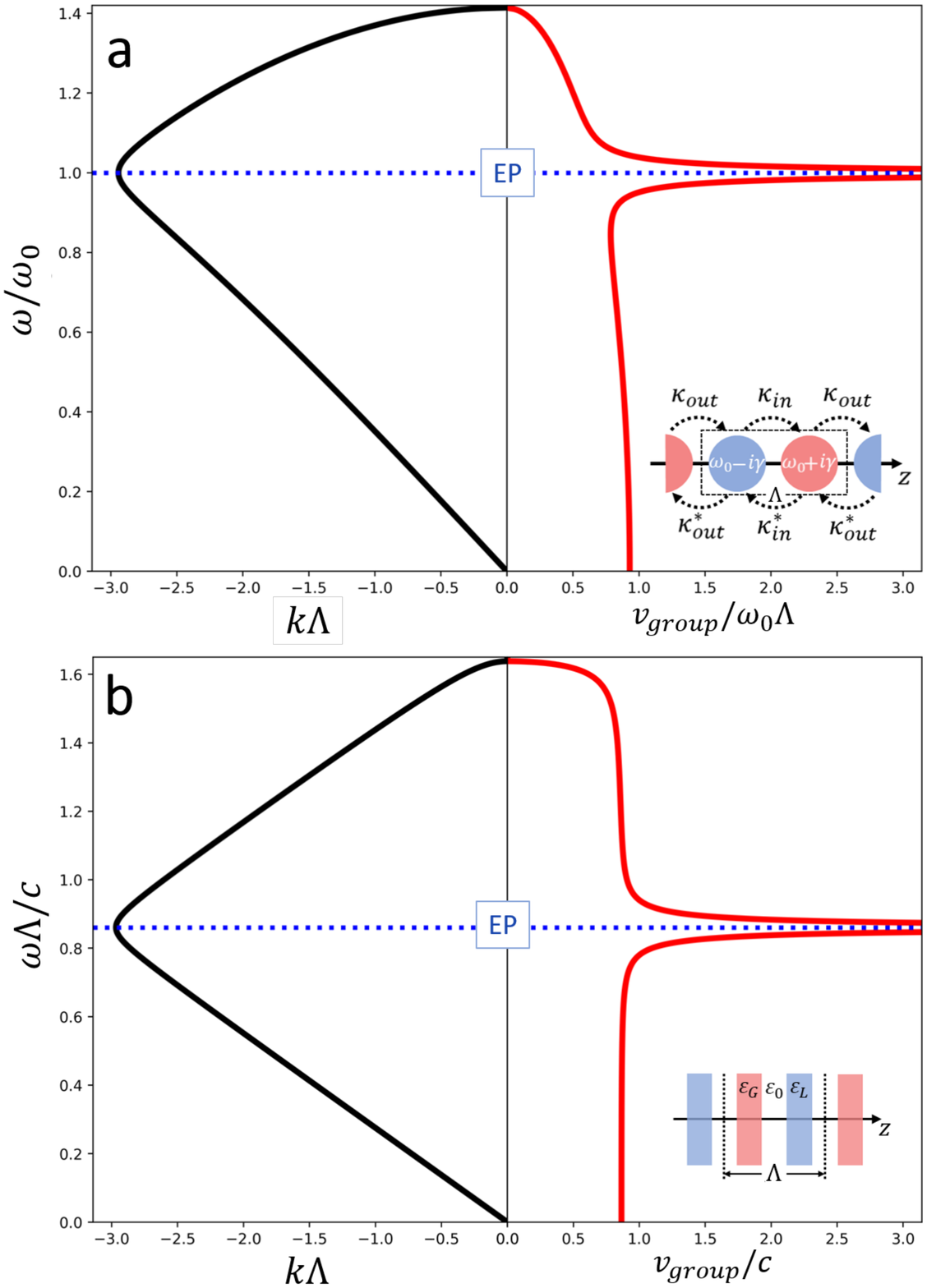}
    \caption{BS (left, black solid line) and group velocity (right, red solid line) of (a) a $\mathcal{PT}$-symmetric SSH chain and (b) a $\mathcal{PT}$-symmetric Bragg stack. The inset in each figure sketches the corresponding system. Dashed blue lines indicate the positions of EPs of the system. The SSH parameters are $\gamma/\omega_0 = 0.1$ and $\kappa_\text{in} = \kappa_\text{out} = \frac{1}{2}\sqrt{\omega_0^2 + \gamma^2} $, while $\varepsilon_\text{G/L} = 1.7 \pm  0.5\imag$ and $\varepsilon_0=1$ for the Bragg stack.
    }
    \label{fig:figure_1}
\end{figure}

In the case of the $\mathcal{PT}$-symmetric SSH model, the complex nature of the BS becomes crucial for understanding the system's behavior.
Conceptually, the complex BS is, of course, a complex multi-valued function of a complex argument ($\{\omega, k\} \in \mathbb{C}^2$). Admittedly, such a 4-dimensional picture is hard to comprehend, visualize and analyze. On the other hand, this space also contains different planes, that represent simplified situations relevant to physical scenarios. As an example, the case of the real-valued frequency cut ($\omega \in \mathbb{R}$ and $k \in \mathbb{C}$) is relevant to systems being harmonically driven at a well-defined frequency, e.g., a narrow continuous-wave (CW) laser, while a real-valued wave vector projection ($k \in \mathbb{R}$ and $\omega \in \mathbb{C}$) would seem relevant to the temporal decay dynamics following an initially well-defined spatial composition of the field.   
Therefore, depending on the physical situation of interest, we have to choose between real-$\omega$-complex-$k$ and complex-$\omega$-real-$k$ representations and calculate the BS with different methods. 
In the former case, we may first find a transfer matrix of the system and then diagonalize it.
The BS obtained with this method is presented in~Fig.~\ref{fig:figure_1}(a). 
In the latter case, we may obtain the BS by diagonalizing the Hamiltonian of Eq.~(\ref{eq:sshHamiltonian}) itself.
This way is easy and results in a simple equation for the BS (assuming, without loss of generality, $\kappa_1$ and $\kappa_2$ to be real),
\begin{equation}\label{eq:SSHexpliciteBS}
 \omega(k) = \omega_0 \pm \sqrt{\kappa_\text{in}^2 + \kappa_\text{out}^2 + 2\kappa_\text{in}\kappa_\text{out}\cos(k \Lambda) - \gamma^2},
\end{equation}
where $\Lambda$ is the length of the diatomic unit cell.

Independently of the representation, one can observe that the SSH chain described by Eq.~(\ref{eq:sshEquation}) with the Hamiltonian of Eq.~(\ref{eq:sshHamiltonian}) has BS properties similar to the ones described in Sec.~II. 
Indeed, one of the main properties of $\mathcal{PT}$-symmetric Hamiltonians is that they have real-valued eigenenergies in the unbroken phase, which produce smooth back-bending lines of real frequencies and complex loops with EPs between them. The
EP is a point in parameter space where the eigenvalues are degenerate and the eigenstates coalesce, and the system has its abrupt phase transition. 
Since the BS has real-valued lines, the BS derivative $\partial \omega / \partial k$ for $\omega \in \mathbb{R}$ gives us the group velocity, and for the 1D problem at hand, the inverse of that is proportional to the DOS~\footnote{In the spirit of electronic band structures, this readily follows from a projection of all the points in the $\omega$-$k$ space onto the $\omega$-axis. As we show rigorously in Appendix~\ref{appendix:DOS}, this causes a vanishing DOS at the EP, while quantum electrodynamics at the EP can still be of interest~\cite{Kanbekyan:2020}.}. 
But just as in the case of dispersive PhCs, this BS implies superluminal dynamics, with even infinite group velocities as one approaches the EP. This obviously unphysical result raises a curiosity for the assumptions and approximations responsible for this artifact of the theory.

\begin{figure*}
    \centering
    \includegraphics[width=\textwidth]{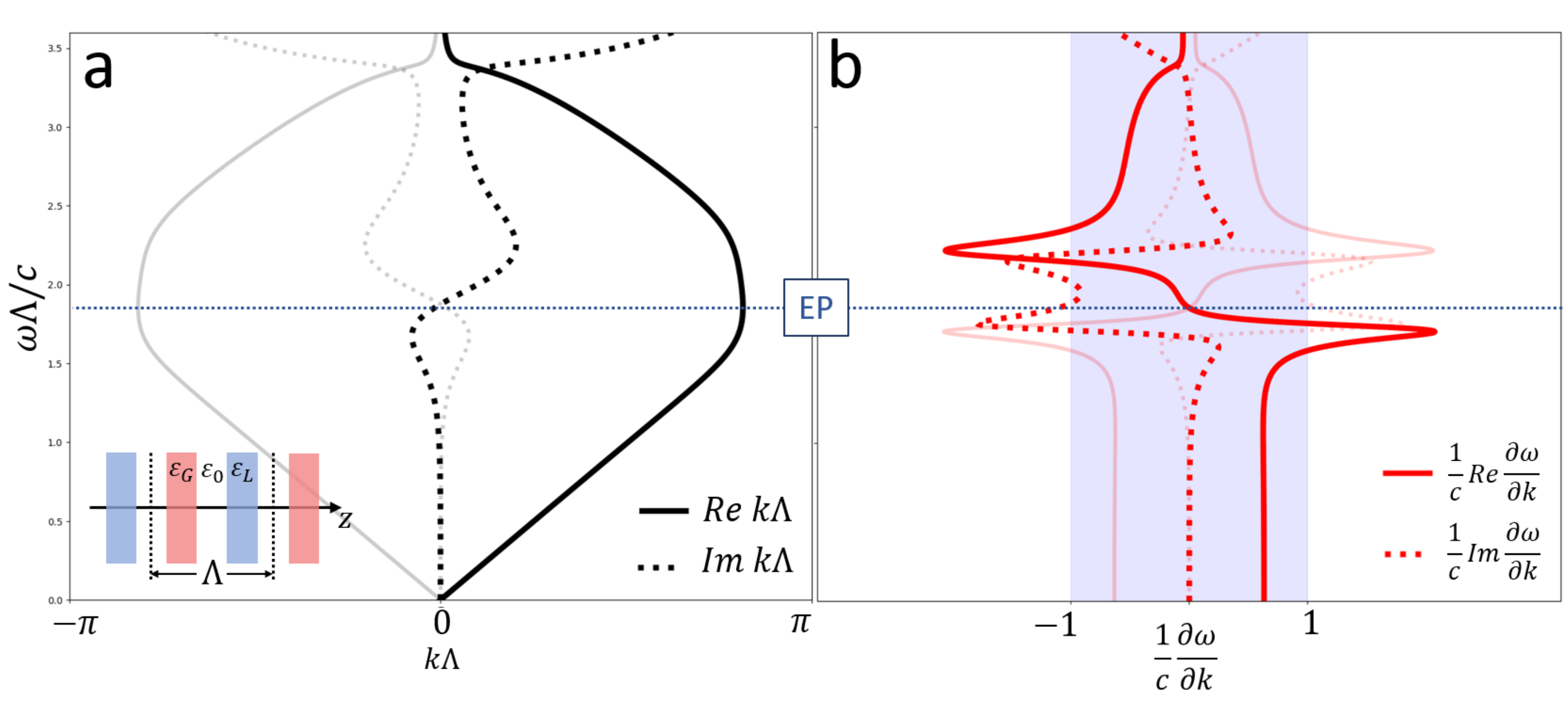}
    \caption{(a) BS and (b) BS derivative of a $\mathcal{PT}$-symmetric Bragg stack with dispersive materials. A sketch of the system is shown at the inset of (a). Solid lines depict the real parts and dashed lines show the imaginary parts of the BS and its derivative. The blue-shaded region in (b) highlights the subluminal regime. The parameters are $\omega_0\tau=3.226$, $\alpha/\omega_0 = 0.6$, and $\varepsilon_b = 4$.}
    \label{fig:band_structure_disp_Bragg}
\end{figure*}

\subsection{Infinite group velocity in Bragg stacks}
A first natural hypothesis which one could suggest is that the SSH model is oversimplified and inadequate for $\mathcal{PT}$-symmetric systems. 
The photonic tight-binding model presented here is derived within CMT by discretizing the exact partial-differential equation, finding modes for each individual element of the discretized system, and coupling parameters between them. 
As a result, we approximate the behavior of the wave equation by the hopping between any lattice site with its increasingly distant neighbors~\cite{Busch:2011}; the tight binding model emerges by just retaining the dominant coupling, which is to the next neighbors.
Assuming that the physically incorrect dynamics is rooted in these simplifications, it would disappear once we included more terms --- in principle terms to all orders --- in our series expansion.
Another aspect that might suggest a tight-binding model to be oversimplified is the fact that the speed of light $c$ does not appear anywhere in it and therefore the model has no way to ``know'' which group velocities might be unphysical.

Therefore, we test this first hypothesis by solving a Bragg-stack modelled after the tight-binding chain, i.e. a 1D $\mathcal{PT}$-symmetric photonic crystal, composed of weakly coupled slab resonators with gain and loss. 
This toy model mimics the initial SSH chain while rigorously retaining the full electrodynamics governed by the Helmholtz equation for the electric field $E$, 
\begin{equation}\label{eq:HelmholzEq}
    \frac{d^2}{dz^2}E + \frac{\omega^2}{c^2}\varepsilon(z)E = 0,
\end{equation}
where $\varepsilon(z)$ is the periodic permittivity along the stacking direction $z$. 
The system is shown in the inset of Fig.~\ref{fig:figure_1}(b) and consists of periodically arranged slabs with gain and loss imitating the corresponding SSH chain sites, separated by a vacuum.
Compared to the above SSH problem leading to Eq.~(\ref{eq:SSHexpliciteBS}), the formulation inherently includes inter and intra-cell couplings to all orders, while $c$ --- the speed of light in vacuum --- is also explicitly present in this formulation.
In the spirit of the SHH model, we consider propagation along the $z$-axis in a positive direction, while $\Lambda$ is the length of the unit cell.

To calculate the BS of such a a system, governed by the exact Helmholtz equation, we use the scattering-matrix method~\cite{Botten:2001}, and thus obtain the BS in the real-$\omega$--complex-$k$ representation.
 Evidently, this BS in Fig.~\ref{fig:figure_1}(b) exhibits qualitatively the same behavior as the one emerging from the corresponding SSH formulation, Fig.~\ref{fig:figure_1}(a): it has the same back-bending, and the group velocity reaches superluminal values and diverges at the EP. 
 These findings refute our initial suggestion that the explanation should be observed beyond short-range approximations inherent to the SSH model.
 
\subsection{Origin of unphysical pulses: broken causality}
As a precursor for the answer
to the problem, we return to the case of the metallic PhCs. 
We have already mentioned that lossless models of metallic PhCs predict back-bendings in the BS conceptually similar to the ones we have for $\mathcal{PT}$-symmetric photonic systems, leading to divergent group velocities.
Such an unphysical behavior results from neglecting material losses. While this approximation might initially seem fairly innocent, Kramers-Kronig relations are formally being violated no matter how low the loss is, and the system is left non-causal. In particular, this is most clearly exhibited in regions with strong dispersion where the group velocity deviates significantly from $c$.
Restoring causality, by including frequency-dispersive loss in accordance with Kramers-Kronig relations into the model, solves the problem of instant pulse propagation.
In a similar way, we hypothesize that the $\mathcal{PT}$-symmetric photonic systems discussed above also violate Kramers-Kronig relations, while exploiting a causal permittivity into the model would restore proper physical dynamics of the system.

This suggestion has foundations stronger than just an analogy with metallic PhCs. 
It was explicitly shown by Zyablovsky \emph{et al.}~\cite{Zyablovsky:2014} that $\mathcal{PT}$-symmetric photonic systems violate the Kramers-Kronig relations if MD is neglected, which renders the common dispersionless model non-causal and thus fundamentally incorrect. Unfortunately, the incorrectness is strongly exhibited in the vicinity of the exceptional point. 
Furthermore, Ref.~\cite{Zyablovsky:2014} shows that systems satisfying Kramers-Kronig relations can only be $\mathcal{PT}$-symmetric for a discrete set of frequencies of the incoming pulse.

However, despite these important findings, Ref.~\cite{Zyablovsky:2014} has remained underappreciated in the rapidly evolving community centered around the physics of $\mathcal{PT}$-symmetric systems and their EPs. 
On the other hand, even when pragmatically neglecting the dispersive properties of constituents in SSH descriptions, the simplified model has successfully assisted the qualitative analysis of many experimental observations.
Indeed, there has been an initial attention on implications for experimental realization of $\mathcal{PT}$-symmetric systems and observing EP, while there has been less focus on further fundamental consequences of ignoring MD. 
However, as we explain below, causality-consistent inclusion of MD in a $\mathcal{PT}$-symmetric periodic system's modelling solves the problem of divergent real BS derivatives and instant pulse propagation in the vicinity of the EP. 

To be more concrete and aid transparent analysis, we invoke a generic Lorentzian-shaped MD profile into the model:
\begin{equation}\label{eq:LorenzMD}
    \varepsilon_\text{G/L}(\omega) = \varepsilon_b \pm \frac{\alpha}{\omega - \omega_0 + \imag\tau^{-1}},
\end{equation}
where $\varepsilon_{b}$ is the background static permittivity, $\alpha$ represents the eigenfrequency $\omega_{0}$ multiplied by an appropriate oscillator strength, and $\tau^{-1}$ governs the spectral width.
By construction, this inherently satisfies the Kramers-Kronig relations,
while allowing us to retain a simple Bragg-stack model.
We immediately see that any system with this MD would satisfy the $\mathcal{PT}$-symmetry only for the particular frequency $\omega = \omega_0$, while $\mathcal{PT}$-symmetry is formally broken away from this frequency. 

In the $\mathcal{PT}$-symmetric Bragg stack, the full MD can be introduced simply by replacing the permittivities with their dispersive counterparts. 
The way to consistently include MD in the SSH chain is perhaps less straightforward.
From the point of view of solid-state physics, tight-binding models are  ultimately a description of a lattice formulated in terms of a (truncated)  Wannier basis and the common CMT is analogous to the linear combination of atomic orbitals in semiconductor physics.
Therefore, we can insert the eigenmodes $|\mathbf{E}_n(\mathbf{r})\rangle$ of the individual nondispersive 
optical resonators~\footnote{We note that, strictly speaking, this only works for bound states, e.g. in waveguide arrays. A rigorous description for coupled open resonators based on quasinormal modes (see e.g. Ref.~\cite{Kristensen:2020}) would be beyond the scope of this paper.} as approximate Wannier functions in the expressions 
in Ref.~\cite{Busch:2011}.
Within the appropriate approximations (see 
Appendix~\ref{appx:dispresive_cmt}) this leads to an eigenvalue problem of the same form as Eq.~\eqref{eq:sshHamiltonian} except with dispersive gain/loss terms 
\begin{align}\label{eq:dispersiveGamma}
\gamma \rightarrow \gamma[1 - \imag (\omega - \omega_0)\tau]^{-1},
\end{align}
which of course can no longer be purely imaginary due to causality.
In order to calculate the BS for both the Bragg stack and the SSH systems, taking MD rigorously into account, 
we must next solve an implicit eigenvalue problem by calculating the complex-valued Bloch wavenumber as a function of frequency. 
To do this, in the case of the Bragg stack we use the scattering-matrix method~\cite{Botten:2001}, while in the case of the SSH chain we use the transfer-matrix method~\cite{Lambropolus:2019}.

\begin{figure}
    \centering
    \includegraphics[width=9cm]{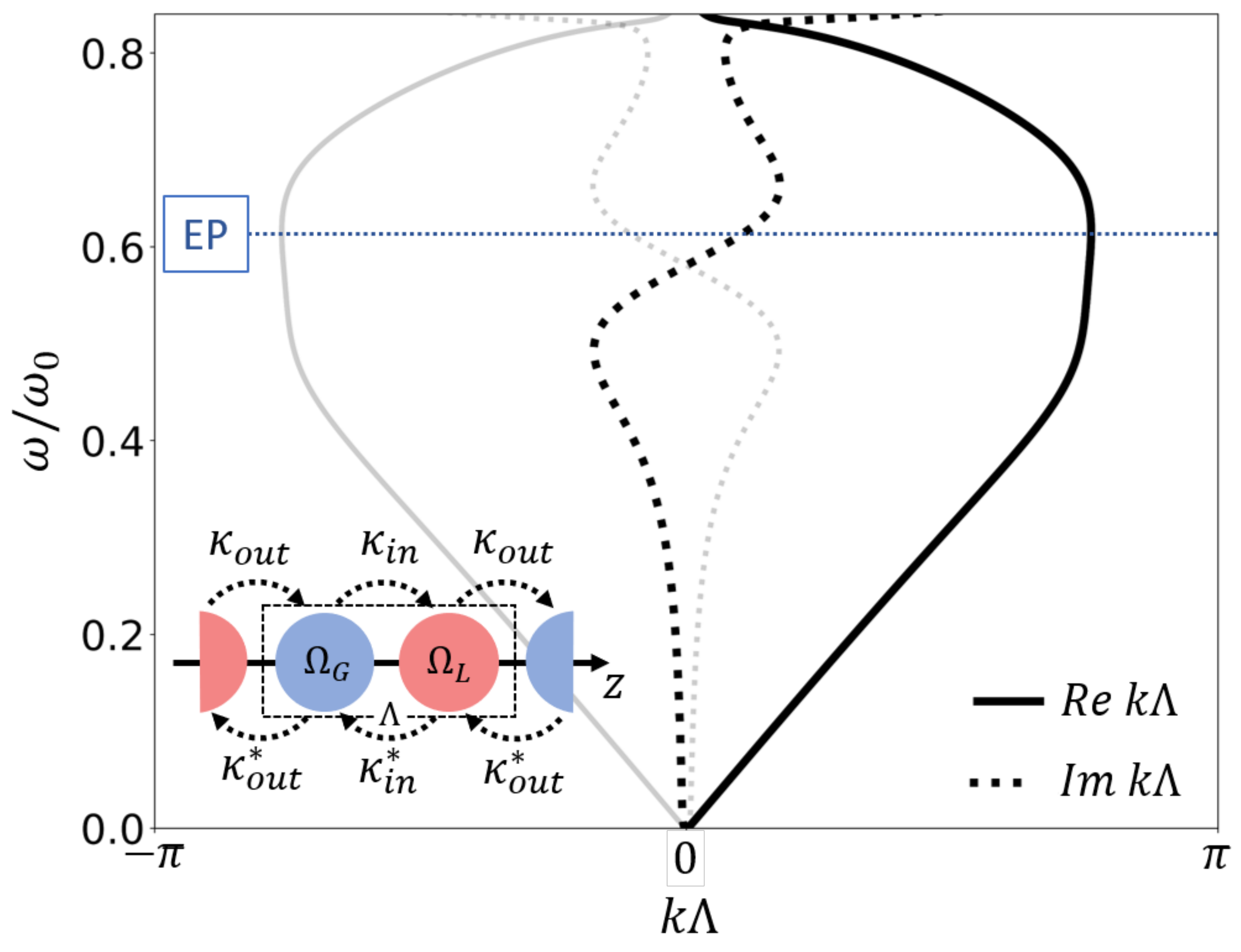}
    \caption{Band structure of a dispersive $\mathcal{PT}$-symmetric SSH chain. A sketch of the system is shown in the inset, where $\omega_\text{G/L} = \omega_0 \pm \gamma[1 - \imag (\omega -\omega_0)\tau]^{-1}$ are dispersive on-site elements according to Eq.~(\ref{eq:dispersiveGamma}). Solid lines show the real parts while dashed lines depict the imaginary parts of the BS. The parameters are the same as in Fig.~\ref{fig:band_structure_disp_Bragg}.
    }
    \label{fig:band_structure_disp_SSH}
\end{figure}

Figure~\ref{fig:band_structure_disp_Bragg} shows the BS and the BS derivative of the dispersive $\mathcal{PT}$-symmetric Bragg stack, and
Fig.~\ref{fig:band_structure_disp_SSH} shows the BS of the dispersive $\mathcal{PT}$-symmetric SSH chain.
One can see that the two BSs are qualitatively the same, and thus the optical properties of the two systems are similar.
We emphasize that the observed dynamics is now in full accordance with the principle of causality, and the apparent violation of the common limits of group velocities to subluminal speeds does not manifest over extended propagation lengths, since the relevant frequency regimes are now associated with also imaginary wave-vector components. 
Physically, the imaginary part of the Bloch wavenumber signifies a decaying wavepackage, its magnitude being exactly such that it prohibits transfer of information at superluminal speeds over finite distances. 
This interplay of dispersion and propagation distance has a parallel in the loss-limited group delay that one faces in passive slow-light waveguides~\cite{Pedersen:2008,Grgic:2012,Raza:2010,Grgic:2011}.

\section{Discussion}
By restoring causality and introducing MD into the model we have fixed the unphysical predictions of gain-loss balance based $\mathcal{PT}$-symmetric systems.
However, this modification fundamentally affects the model and could potentially distort predictions derived within the non-dispersive approximation. 
For instance, the discrete set of frequencies allowing $\mathcal{PT}$-symmetry discussed in Ref.~\cite{Zyablovsky:2014} constitutes such a fundamental change. 

Another example is related to the elimination of the infinite BS derivatives at the EP.
In the non-dispersive model in the vicinity of an EP, the BS obeys a square root law, linking frequency not only with the Bloch phase, but also with the system's inherent parameters, which is evident in the case of the tight-binding chain from Eq.~(\ref{eq:SSHexpliciteBS}). 
Thus, eigenfrequencies as a function of loss and coupling parameters have the same behavior, bifurcating at the EP and varying with this parameter as a square root. 
This implies that eigenfrequencies of a perturbed system at the EP split, and the magnitude of this split is proportional to the square root of the perturbation~\cite{Pick:2017}.
This effect formed the basis for ideas of sensitivity enhancement at EP since such square-root dependence is steeper then a linear one~\cite{Wiersig:2014}, theoretically allowing to detect smaller perturbations~\cite{Wiersig:2020}. 
However, inclusion of MD changes the picture; 
it adds imaginary slope to eigenfrequencies, which fixes the problem of instantaneous propagation, but importantly it also weakens the above-mentioned square-root response. 
This becomes self-evident from comparison of the BS at the EP of dispersive and non-dispersive systems in Fig.~\ref{fig:figure_1}(b) and Fig.~\ref{fig:band_structure_disp_Bragg}(a), respectively.
The same happens to the bifurcation of eigenfrequencies as functions of loss parameter since it is directly related to the BS.
As such, limitations due to causality and frequency dispersion may become another challenge for the prospects of EP-based sensitivity enhancement. 

We emphasize that this change is fundamental and occurs with any amount of dispersion to be introduced.
The system is highly responsive for low dispersion as the imaginary slope at EP is proportional to the spectral width $\tau^{-1}$, thus the weaker the MD, the steeper this slope.
However, approaching this limit requires ever increasing oscillator strengths $\alpha$, since the $\mathcal{PT}$-symmetry parameter is $\gamma = \alpha \tau$ on resonance (see Appendix~\ref{appendix:BSderivative}).
The above discussion illustrates that retaining $\mathcal{PT}$-symmetric systems causal fundamentally changes the entire picture and may affect many drawn conclusions and features derived from non-dispersive models.  

\section{Conclusions and Outlook}

We studied propagation of light in $\mathcal{PT}$-symmetric photonic systems based on balanced gain and loss. 
We considered $\mathcal{PT}$-symmetric SSH photonic chains and Bragg stacks and pointed out unphysical predictions, following from the non-dispersive material description:
such systems allow infinite purely real group velocities at the EP, which leads to instantaneous light propagation.
We argued that this is the result of the non-causal nature of broadband $\mathcal{PT}$-symmetry based on gain-loss balance, while MD should be included consistently into the model
We showed that causal $\mathcal{PT}$-symmetric systems now have complex BSs and BS derivatives with imaginary slope at the EP, which eliminates the issue of otherwise instantaneous prorogation.
Finally, we discussed the fundamental impact of MD on other features, showing that dispersive systems do not exhibit the anticipated square-root dependence of $\omega$ on $k$ and system parameters, thus affecting the realistic performance of sensing devices based on $\mathcal{PT}$ symmetry.

\section{Acknowledgements}

C.~W. acknowledges funding from a MULTIPLY fellowship under the Marie Sk\l{}odowska-Curie COFUND Action (grant agreement No. 713694).
N.~A.~M. is a VILLUM Investigator supported by VILLUM FONDEN (grant No. 16498).

\begin{appendix}

\section{Dispersive CMT model}
\label{appx:dispresive_cmt}

As mentioned in the main text, we use the magnetic field of the
eigenmodes $H_n(\mathbf{r})$ in the $n$-th resonator as approximate 
Wannier functions in the expressions for a 1D $H$-polarized problem in 
Ref.~\cite{Busch:2011}.
Using $E_n = \imag \omega_0 \partial_x H_n$, we
obtain the nonlinear (self-consistent) eigenvalue problem
\begin{align}
\sum_n
\Big[ \omega_0^2 \langle E_{n'} | \varepsilon^{-1}(\omega)  | E_{n} \rangle
- \omega^2 \underbrace{\langle H_{n'} | H_{n} \rangle}_{\approx \mathbb{I}} \Big] a_n = 0,
\end{align}
where $\langle H_{n'} | H_{n} \rangle \approx \mathbb{I}$ is the
eigenmode normalization.
Next, we assume $\varepsilon(\omega) = \varepsilon_b + \delta\varepsilon(\omega)$ with
\begin{align}
\delta\varepsilon(\omega) 
= \frac{\imag \alpha \tau}{1 - \imag (\omega - \omega_0)\tau}
\ll \varepsilon_b,
\end{align}
which allows us to approximate
\begin{align}
\langle E_{n'} | \varepsilon^{-1}(\omega)  | E_{n} \rangle
=
\underbrace{ \langle E_{n'} | \varepsilon^{-1}_b  | E_{n} \rangle}_{\mathcal{C}_{n'n}}
- \underbrace{
\langle E_{n'} | \varepsilon_b^{-2} \delta\varepsilon(\omega)  | E_{n} \rangle}_{\mathcal{D}_{n'n}},
\end{align}
without breaking causality.
We only consider terms with $|n'-n| \leq 1$ (next-neighbour coupling).
In case of an alternating gain-loss pattern as required for $\mathcal{PT}$-symmetry, $\mathcal{D}_{n'n} = 0$ for odd $|n'-n|$, so we obtain 
a matrix eigenvalue problem $\mathcal{M} \mathbf{a} = \omega^2 \mathbf{a}$ where $\mathcal{M}$ 
has the same form as $\mathcal{H}$ in Eq.~\eqref{eq:sshHamiltonian}.
Under the additional assumption $|\omega - \omega_0| \ll \omega_0$, this
can be further linearized by Taylor-expanding the eigenfrequency to find
$ \mathcal{H} \mathbf{a} = \omega \mathbf{a}$ as in Eq.~\eqref{eq:sshHamiltonian} with
\begin{subequations}
\begin{align}
\kappa_\text{in} = & \frac{\omega_0}{2}
\langle E_{0} | \varepsilon_b^{-1} | E_{1} \rangle,
\quad
\kappa_\text{out} = \frac{\omega_0}{2}
\langle E_{1} | \varepsilon_b^{-1} | E_{2} \rangle,
\\
\gamma = & \frac{\omega_0}{2}
\langle E_{0} | \varepsilon_b^{-2} \delta\varepsilon(\omega)  | E_{0} \rangle \simeq \frac{\alpha \tau}{1 - \imag  (\omega - \omega_0)\tau}.
\end{align}
\end{subequations}
This means that, to lowest order, we recover the same type of Hamiltonian
except for the transformation already given in Eq.~\eqref{eq:dispersiveGamma}
if we identify $\alpha \tau$ with the original nondispersive $\gamma$ for
compatibility  in the nondispersive limit $\tau \rightarrow 0$.

\section{BS derivative at a dispersive EP} 
\label{appendix:BSderivative}
Here we calculate how the inclusion of inevitable material dispersion reduces
the square root law of a real EP to a linear BS derivative.
To this end, we start with the upper branch of Eq.~\eqref{eq:SSHexpliciteBS} 
(the lower branch behaves analogously) and transform 
$\gamma \rightarrow \gamma (1 - \imag \tau \Delta)^{-1}$ according to 
Eq.~\eqref{eq:dispersiveGamma}, 
\begin{align}
\underbrace{
\Delta + \sqrt{\kappa_\text{in}^2 + \kappa_\text{out}^2 
+ 2 \kappa_\text{in}  \kappa_\text{out} \cos k \Lambda 
- \frac{\gamma^2}{(1 - \imag  \Delta\tau)^{2}}}
}_{F(\Delta, k)} = 0,
\end{align}
where we also introduced the frequency detuning 
$\Delta = \omega - \omega_0$.
This defines the BS implicitly by the equation $F(\Delta, k) = 0$
and therefore we can find the BS derivative via the implicit-function theorem:
\begin{align}
\frac{\partial \omega}{\partial k} = \frac{\partial \Delta}{\partial k}
= -\frac{\partial_k F(\Delta, k)}{\partial_\Delta F(\Delta, k)}.
\end{align}
We find:
\begin{subequations}
\begin{align}
\partial_\Delta F(\Delta, k) = &
1 + \frac{\imag \tau \gamma^2}{\sqrt{\ldots} (1 - \imag  \Delta\tau)^3},
\\
\partial_k F(\Delta, k) = & \frac{-1}{\sqrt{\ldots}} \kappa_\text{in} \kappa_\text{out} \Lambda \sin k \Lambda,
\\
\Rightarrow \quad
\frac{\partial \Delta}{\partial k} = & \frac{\kappa_\text{in} \kappa_\text{out} \Lambda \sin k \Lambda}{\sqrt{\ldots} + \imag \tau \gamma^2 (1 - \imag  \Delta \tau)^{-3}}, 
\end{align}
\end{subequations}
where $\sqrt{\ldots}$ is the square root appearing the the definition of 
$F(\Delta, k)$.
At the EP, we have: $\sqrt{\ldots} = 0$ and $\Delta = 0$,
so the band structure derivative at the EP is given as:
\begin{align}
\frac{\partial \omega}{\partial k} \Bigg|_{\text{EP}} =
- \imag \frac{\kappa_\text{in} \kappa_\text{out} \Lambda \sin k \Lambda}{\tau \gamma^2}.
\end{align}
This means that the BS slope is inversely proportional to the 
spectral width of the gain/loss resonance.

\section{Total DOS}
\label{appendix:DOS}
Here, we quickly investigate the \emph{total} DOS of a 
$\mathcal{PT}$-symmetric system.
We assume the Hamiltonian to commute with the combined $\mathcal{PT}$ operator: 
$(\mathcal{PT}) \mathcal{H} (\mathcal{PT})^{-1} = \mathcal{H}$.
If the Hamiltonian commutes with this operator, then so does its (the Hamiltonian's) inverse, i.e. the Green operator ${\mathcal{G}} = \mathcal{H}^{-1}$:
\begin{align}
(\mathcal{PT}) {\mathcal{G}} (\mathcal{PT})^{-1} = {\mathcal{G}}.
\end{align}
In practice, $\mathcal{G}$ takes the form of an integral form whose 
kernel we call the Green function. 
This $\mathcal{PT}$ symmetry translates to the Green function:
\begin{align}
{G}(x, x'; t) = G(-x, -x'; -t),
\end{align}
for all $x$, $x'$ and $t$.
Furthermore, in classical physics the Green function in real space and
time-domain is real valued.

We can now formulate the total DOS at frequency $\omega$:
\begin{align}
\rho(\omega) = C \int_\Omega \total x \ \Im\{ {G}(x, x; \omega)\},
\end{align}
where $C$ includes all necessary prefactors and
$\Omega$ is the domain of the Hamiltonian (e.g. a unit cell in a 
periodic system).
We insert the Fourier transform defining ${G}(x, x; \omega)$ to find:
\begin{subequations}
\begin{align}
\rho(\omega) = & \frac{C}{2\pi} \Im \Big\{ 
\int_\Omega \total x \int_{-\infty}^\infty \total t \ {G}(x, x; t) \exp(\imag \omega t) \Big\}
\\
 = & \frac{C}{2\pi}  \int_{-\infty}^\infty \total t \ 
 \sin(\omega t) \underbrace{\int_\Omega \total x \ {G}(x, x; t)}_{\overbar{G}(t)} .
 \end{align}
\end{subequations}
We can next use the $\mathcal{PT}$ symmetry of $\mathcal{G}$ to find:
\begin{subequations}
\begin{align}
 \overbar{G}(t) = & \int_{\Omega^+} \total x \ [
 {G}(x, x; t) + {G}(-x, -x; t) ]
 \\
 = & \int_{\Omega^+} \total x \ [
 {G}(x, x; t) + G(x, x; -t) ],
\end{align}
\end{subequations}
where $\Omega^+$ is the half of $\Omega$ where $x>0$ [more generally any 
domain with $\Omega^+ \cup \mathcal{P} \Omega^+ = \Omega$].
So, $ \overbar{G}(t) = \overbar{G}(-t) $  turns out to be an even function. 
Therefore the overlap with $\sin(\omega t)$ vanishes and $\rho(\omega) = 0$.
Of course, this does not say anything about the \emph{local} DOS $\rho(x, \omega)$ except that it should be odd with respect to $x$, 
i.e. $\rho(-x, \omega) = -\rho(x, \omega)$.

\end{appendix}

\bibliography{references}

\end{document}